%% file: main.tex
\documentclass[conference]{IEEEtran}[en]
\usepackage{cite}
\usepackage{amsmath,amssymb,amsfonts}
\usepackage{algorithmic}
\usepackage{graphicx}
\usepackage{textcomp}
\def\BibTeX{{\rm B\kern-.05em{\sc i\kern-.025em b}\kern-.08em
    \documentclass{article}
T\kern-.1667em\lower.7ex\hbox{E}\kern-.125emX}}
\usepackage{graphicx}
\usepackage{array}
\usepackage{listings}
\usepackage{url}
\usepackage{subfig}
\usepackage{caption}
\usepackage{listings}
\usepackage{placeins}
\usepackage{multirow}
\usepackage[nounderscore]{syntax}
\usepackage{lipsum}% http://ctan.org/pkg/lipsum
\usepackage{multicol}% http://ctan.org/pkg/multicols
\usepackage{graphicx}% http://ctan.org/pkg/graphicx
\usepackage{hhline}
\usepackage{color}
\usepackage{makecell}
\usepackage{xcolor}
\usepackage{gensymb}
\usepackage{subfloat}
\usepackage{tabularx}
\usepackage{comment}
\usepackage{graphicx}
\usepackage{listings}

\newcommand{\ie}{\textit{i}.\textit{e}.,\ }
\newcommand{\eg}{\textit{e}.\textit{g}.,\ }

\newcommand{\1}{\textit{(a)}}
\newcommand{\2}{\textit{(b)}}
\newcommand{\3}{\textit{(c)}}
\newcommand{\4}{\textit{(d)}}

\newsavebox{\mybox}

\begin{document}

\title{SECAdvisor: a Tool for Cybersecurity Planning using Economic Models}

\author{\IEEEauthorblockN{
        Muriel Figueredo Franco, Christian Omlin, Oliver Kamer, Eder John Scheid, Burkhard Stiller}

\IEEEauthorblockA{
    Communication Systems Group CSG, 	Department of Informatics IfI, University of Zurich UZH\\
	Binzm{\"u}hlestrasse 14, CH--8050 Z{\"u}rich, Switzerland\\
	\\
	E-mail: \{franco, scheid, stiller\}@ifi.uzh.ch, \{christian.omlin, oliverluca.kamer\}@uzh.ch
}
}

\maketitle

\begin{abstract}
Cybersecurity planning is challenging for digitized companies that want adequate protection without overspending money. Currently, the lack of investments and perverse economic incentives are the root cause of cyberattacks, which results in several economic impacts on companies worldwide. Therefore, cybersecurity planning has to consider technical and economic dimensions to help companies achieve a better cybersecurity strategy. This article introduces SECAdvisor, a tool to support cybersecurity planning using economic models. SECAdvisor allows to \1 understand the risks and valuation of different businesses' information, \2 calculate the optimal investment in cybersecurity for a company, \3 receive a recommendation of protections based on the budget available and demands, and \4 compare protection solutions in terms of cost-efficiency. Furthermore, evaluations on usability and real-world training activities performed using SECAdvisor are discussed. 

\end{abstract}

\section{Introduction}
One challenge for cybersecurity is how to plan a cybersecurity strategy without overspending money on protection measures \cite{CyberTEA}. The cybersecurity market is worth billions of dollars and steadily rising investments, with companies investing in cybersecurity to ensure availability and protect their core businesses from economic losses. These losses might include direct losses due to business interruption (\eg an e-commerce that cannot offer products due to server's downtime) or indirect losses like reputation harm and legal penalties.

Although there are businesses more prone to specific attacks, in general, attackers tend not to spend too much time focusing on one specific company but on exploring vulnerabilities in the type of businesses they see as potential weaknesses. This happens especially in the case of Small and Medium-sized Enterprises (SMEs), which are the focus of more general attacks (\ie not tailored for a specific company) \cite{ENISA-Survey} because attackers know the reality of most SMEs: lack of training, limited technical expertise, and insufficient budget for cybersecurity.

%\cite{SecGrid, saci} 
Cyberattacks can devastate SMEs and put many companies out of the market in the last few years. Therefore, It is important to understand and mitigate risks to reduce possible impacts on their operations \cite{SecGrid, saci}. However, most of these companies do not have sufficient budgets to spend, making cybersecurity planning harder. Therefore, it is important to have tools and models that simplify the task of cybersecurity planning, making it not only more user-friendly but also more cost-efficient. Thus, approaches that rely on cybersecurity economics concepts have to be considered \cite{Economics-ENISA} together with technical knowledge to achieve a balance between risks and investments for a company.

In cybersecurity economics, the Gordon-Loeb (GL) model is the most well-accepted analytical model to determine the optimal investment level in cybersecurity \cite{gordon_2002}. The model considers \1 how much the data or service is valued, \2 how much the data is at risk (\eg attack probability-based historical data), and \3 the probability that an attack is going to be successful, which can be defined based on the threat modeling and risk analysis. Also, extensions to the GL model have been proposed over the years. 

In the last years, information segmentation was also introduced as a key element for investments in cybersecurity \cite{gordon-segmentation}. The information segmentation argues that the amount invested in cybersecurity, when calculated using the GL model, should be considered in terms of specific information segment and their potential benefits (\ie investments vs. potential losses). However, this kind of model is not trivial to be applied by companies, nor is it well-known by non-technical users. Therefore, solutions that support the application of GL and other economic metrics to cybersecurity are welcome for companies' faster adoption since economic motivation is one of the strongest to convince a company to invest in~cybersecurity.

Based on that, SECAdvisor, an open-source and visual tool for calculating the optimal investment in cybersecurity, is proposed. SECAdvisor is the first of its kind. It allows users to define information segments within a company and calculate the optimal investment for each segment, including potential losses with and without an optimal investment in cybersecurity. This calculation applies GL to estimate values accurately. After calculating optimal investments in cybersecurity, SECAdvisor can recommend protection measures using an external recommendation engine \cite{MENTOR}. Furthermore, the Return On Security Investment (ROSI) metric is calculated for each recommended solution to compare protections in terms of cost-effectiveness.

The remainder of this paper is organized as follows. Section II introduces key economic models, while Section III reviews related work on cybersecurity economic solutions. Section IV presents the SECAdvisor tool, followed by evaluations as of Section V. Finally, Section V concludes the article.

\section{Background}
This section introduces the theoretical foundations of two of the most well-accepted cybersecurity economics models, including examples for cost analysis and investments in cybersecurity. These models are used as the basis to conduct cybersecurity planning under an economic lens using the SECAdvisor tool.

\subsection{Gordon-Loeb (GL)}
The GL model is an economic model used to analyze the optimal investment level in cybersecurity. The model was proposed in 2002 by Gordon and Loeb \cite{Gordon-Loeb-2002} and takes into account the vulnerability of a system and also the potential financial loss due to a cyberattack. One of the ideas behind the model is that a company should not necessarily invest in mitigating threats in more vulnerable systems. Besides that, a company should focus its efforts on systems with a medium level of vulnerabilities (not on the extreme cases), which is realistic to mitigate risks without substantial financial investments. Also, the analysis conducted by the authors of the model suggests that only a tiny percentage of an expected financial loss due to a cyberattack has to be spent on cybersecurity. Thus, GL provides an economic analysis that provides insights for investment in cybersecurity, but it is still clear that it is not a trivial task, since different elements have to be considered for a precise analysis, such as technical aspects, cyberattacks behaviors, and specific business configurations (\eg sector, maturity, employees, and IT infrastructure).

\begin{figure}[ht]
\begin{center}
    \includegraphics[width=0.50\textwidth]{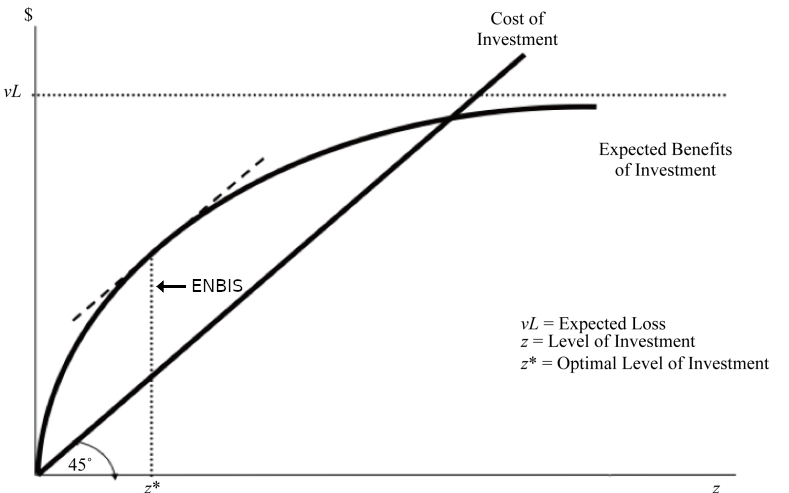}
  \caption{Level of Investment in Cybersecurity from \cite{Gordon-Loeb-2016}}
  \label{fig:gordon}
\end{center}
\end{figure} 

Figure \ref{fig:gordon} shows, based on the investigation conducted in \cite{Gordon-Loeb-2002}, that after a certain threshold, the investments in cybersecurity start to be not worthwhile if compared with the expected loss in case no investments are made. An optimal investment is where the difference between benefits and costs, \ie the Expected Net Benefit of Investment in Information Security (ENBIS), is maximized. Therefore, an investment starts to become good when ENBIS $>$ 0 and the optimal amount invested in cybersecurity is the maximum $ENBIS$ for a given Breach Probability Function (BPF).

The ENBIS is calculated in relation to the Expected Benefit of Investment in Information Security (EBIS) and the amount invested. The EBIS is defined as the reduction in the company's expected loss because of the additional investment made in cybersecurity, which can be described as provided in Equation \ref{eq:ebis}.

\begin{equation}
\label{eq:ebis}
\begin{aligned}
EBIS(z) = [v - S(z, v)]L
\end{aligned}
\end{equation}

Then, the ENBIS can be calculated as the net benefit of the investment made in cybersecurity, which means reducing the expected loss minus the amount invested in cybersecurity. Thus, the ENBIS calculation is shown in Equation \ref{eq:enbis}.

\begin{equation}
\label{eq:enbis}
\begin{aligned}
ENBIS(z) = EBIS - z
\end{aligned}
\end{equation}

A security breach probability function (\ie BPF) is defined as $S(z, v)$, which denotes the probability of a system with a vulnerability $v$ being breached, given that the company has made a cybersecurity investment of $z$. There are two measures used for the productivity of cybersecurity. These measures are determined as $\alpha > 0$ and $\beta \geq 1$.

\cite{Gordon-Loeb-2002} defines two different security breach classes (\ie definitions for $S(z, v)$) to show the performance of the GL model to estimate the optimal investment in cybersecurity. The purpose of a cybersecurity investment is to lower the probability that a system within the company will have a financial loss. Thus, the GL model is proven valid using these two classes of security breach probability functions. The first class refers to a linear vulnerability, while the second analyzed class is concave (\ie the slope of the graph line increases gradually from left to right). It is important to note that, based on the analysis conducted, the optimal investment in cybersecurity is always $\leq \frac{1}{e}$, where $e$ is the Euler's constant (\ie $\approx 2.71828$). This means that the optimal investment is always $\leq 37\%$ of the expected loss (\textit{vL}) without~investments \cite{gordon-37}.

In summary, GL determines, in a general way, that the maximum investment ($z_{max}$) in cybersecurity will never exceed 37\% of the expected loss (\textit{vL}) for all functions part of the classes investigated by \cite{Gordon-Loeb-2002}.
  
In order to calculate the optimal investment, it is needed to use the productivity of a cybersecurity investment, which may vary for different scenarios, depending on specific concerns surrounding a particular set. Another finding from the GL model is that the amount of investment does not always increase based on the level of vulnerability \cite{Gordon-Loeb-2016}. For example, a company can focus more on protecting a system with a medium level of vulnerability than one with a high level of vulnerability. This is a consequence of the productivity of incremental investments in cybersecurity, which means that a given $S(z,v)$ can determine that after or before a certain threshold of vulnerability, investments are not efficient (\eg it will not reduce the chance of a system being attacked or~breached).

In another work, \cite{5438086} provided a counter example for the GL model by constructing a scenario where an investment up to 50\% can be needed. However, \cite{gordon-37} shows that the 37\% rule is valid by proving that security breach functions are not only convex but log-convex (\ie $\log S(z,v)$ is convex for both GL $class\_1$ and $class_2$). Therefore, the GL model is valid for any family of functions that is part of the classes investigated in the original work \cite{Gordon-Loeb-2002}. Thus, as can be seen, the GL has been discussed and improved over the years, making it not perfect but the most well-accepted model for the estimation of cybersecurity investments. However, it is still challenging to precisely determine the optimum investment due to different complexities and nuances involving the cybersecurity domain, such as cybersecurity externalities not mapped in the model when it was proposed and the difficulty of conducting parametric estimations for different real-world scenarios. 

Also, the original GL model has been refined to consider externalities \cite{Gordon-Loeb-Externalities} and the authors provided a new extension to include the concept of information segmentation \cite{gordon-segmentation} during the GL calculation. Besides that, the challenge of calibrating GL parameters (\eg $\alpha$ and $\beta$) for specific real-world scenarios have been considered by different approaches, such as estimating the GL model parameters \cite{gordon-calibrate} and using the GL security breach functions to determine the probability of an insurance claim \cite{gordon-insurance}.

Companies often have several information areas at their disposal, which makes information segmentation inevitable. Segmentation of networks, information, and databases is a practice that facilitates information access to specific individuals while also implementing access control to define who can access specific systems and or information. Different segments cover systems and information with particularities and, consequently, specific values and interests for both company and attackers. For example, a company's financial department might have access to customers' databases and payment systems to handle sensitive information that is very worthy and costly if leaked or compromised. This has to be considered when investing in cybersecurity, since a specific segment might be directly related to the potential benefits of cybersecurity investments.

Thus, based on the GL model, the optimal amount ($z^{*}_{i}$) to invest in a specific information segment $i$ depends on the value of the information ($L_{i})$ that is part of the segment. Also, the vulnerability of each segment ($v_{i}$) has to be considered for the calculation of the productivity of additional investments in cybersecurity for each segment. Therefore, the total cost of investment results in the sum of each segment calculated individually. Hence, it is possible to prioritize segments based on cost-benefits and achieve a better overall cybersecurity investment. In order to find the optimal investment per segment, four steps are required \cite{gordon-segmentation}:

\begin{itemize}
\item \textbf{Step 1}: Estimating the value and therefore the potential loss (\textit{L}) of each segment.
\item \textbf{Step 2}: Estimate the probability ($v$) of each segment's information falling victim to a successful cyberattack.
\item \textbf{Step 3}: Create a grid with all possible combinations of Step 1 and Step 2. Each cell of this grid represents the expected loss (\textit{L}) without cybersecurity investments. The expected loss represents the potential benefit that can be gained by investing in cybersecurity. This means to estimate the productivity of the investments by calculating $S_{i}(z_{i},v_{i})$.
\item \textbf{Step 4}: Derive the level of cybersecurity investment $z^{*}_{i}$ by increasing the investment as long as the benefit of the additional investment is greater than or equal to the cost of the additional investment. Since not all investments in cybersecurity have the same productivity, the optimal amount for investments in different segments will vary.
\end{itemize}

Since each information segment is a subset of the total information, the GL model assumes that the effectiveness of cybersecurity investment in a segment is inversely related to the proportion of the value of the information in a segment given the total company's value of information. Equation \ref{eq:segprob} determines the breach probability function of a segment $i$.

\begin{equation}
\label{eq:segprob}
\begin{aligned}
S_{i}(z_{i},v_{i}) = S(\frac{z_{i}}{\frac{L_{i}}{L}}, v_{i})
\end{aligned}
\end{equation}

The security breach probability function can be defined according to the company. The function used by \cite{gordon-segmentation} is based on an estimation that a Chief Information Officer (CIO) and a CISO of a hypothetical company did together. This is shown in Equation \ref{eq:ciso}. Also, the Equation \ref{eq:optimal} shows how to calculate the optimal investment $z^{*}_{i}$ for a given segment $i$. This means that the optimal investment has to satisfy this equation.

\begin{equation}
\label{eq:ciso}
\begin{aligned}
S(z, v) = \frac{v}{1+\frac{1}{L \times \alpha}\times z}\text{, where } \alpha = 0.001
\end{aligned}
\end{equation}

\begin{equation}
\label{eq:optimal}
\begin{aligned}
S(\frac{z^{*}_{i}}{\frac{L_{i}}{L}}, v_{i}) \times L + 1 = 0
\end{aligned}
\end{equation}

The GL model is mainly considering the information segmentation to assist companies in deriving the investment in cybersecurity with a more cost-benefit and accurate perspective. Companies must to understand that cybersecurity investments are best viewed as a process that focuses on preventing breaches when possible and minimizing the losses from breaches that occur. Therefore, information segmentation can be an efficient approach to help companies better understand how to invest in cybersecurity. Also, it is not realistic to achieve perfect cybersecurity, especially from an economics perspective, even with information segmentation. Thus, part of the cybersecurity investments should be considered to have also an efficient recovery plan in case of cyberattacks surpassing the company's cybersecurity. 

\subsection{Return On Security Investment (ROSI)}
The concept of ROSI is slightly similar to the Return on Investment (ROI). However, while ROI focuses on measuring the benefits/profits made from a particular investment, ROSI focuses on the loss prevented by a cybersecurity investment. ROSI is a cybersecurity economics metric that helps to identify when a given solution (\eg Firewall, Antivirus, or Cybersecurity-as-a-Service product) is cost-efficient or not \cite{ROSI, ENISA-ROSI}. Also, this metric is beneficial when comparing two different solutions with similar characteristics to determine which one should be acquired from an economic perspective. 

A desirable result is ever a ROSI $\geq$ 1, which means that the payback is positive. If ROSI is $\leq$ 1,  there is no cost-benefit in investing in the specific solution. Therefore, the higher ROSI, the better the investment in a solution. ROSI general calculation is provided in Equation \ref{eq:rosi}. As can be seen, for the calculation of ROSI, it is needed to quantify the monetary risk of a cyberattack. Therefore, analytical approaches have to be in place to help companies determine the possible financial losses due to a cyberattack.

\begin{equation}
\label{eq:rosi}
\begin{aligned}
ROSI = \frac{Risk_{Reduction} - Solution_{Cost}}{Solution_{Cost}} \text{, where}\\
Risk_{Reduction} = ALE \times Mitigation_{Ratio}
\end{aligned}
\end{equation}

Besides the solution's cost and efficiency (\ie mitigation rate), ROSI uses the Annual Loss Expectancy (ALE) as input. The calculation of ALE is shown in Equation \ref{eq:ALE}. For that, it is needed to estimate the Annual Rate of Occurrence (ARO) of cybersecurity incidents and also the Single Loss Exposure (SLE), which means that an analysis of the company has to be made in order to understand the history of the attacks to identify its behaviors and impacts in the company.

\begin{equation}
\label{eq:ALE}
\begin{aligned}
ALE = ARO \times SLE, \text{where}\\
\text{ALE: Annual Loss Expectancy}\\
\text{ARO: Annual Rate of Occurrence}\\
\text{SLE: Single Loss Exposure}
\end{aligned}
\end{equation}

SLE can be described as the cost of a loss due to a single incident. As it is the sum of the losses, the value of the loss has to be very objective from company to company. Also, the loss has to include the direct costs (\eg business disruption and recovery) of a cybersecurity incident and indirect costs (\eg reputation and legal impacts).

ARO is the probability of an incident happening in a year. This probability depends on several factors (\eg level of cybersecurity, sector, and market behaviors), and it changes from company to company. If this information is not available within the company, it is possible to use entire sectors (\eg Healthcare, Financial, or Telecom) as a benchmark to support the decision.

For an example of ROSI calculation, suppose an e-commerce company XYZ with an average loss due to cyberattacks of approximately US\$ 30,000, including financial loss due to business interruption, investigation costs, and third-party losses. The past attacks history shows that phishing is the leading cause of the incident in the company, and it strikes roughly three times a year. An anti-phishing product is offered to the company at the price of US\$ 25,000 and the promise to reduce the number of successful attacks by 50\%.

In order to verify if this security product is cost-efficient (\ie the cost of the product is lower than the potential financial loss that the product can avoid), the company uses the ROSI model. ARO, in this case, is equal to 3 (number of times attacks strike the company), and SLE is equal to US\$ 30,000 (average loss due to a single cyberattack). The solution cost is US\$ 25,000, and the mitigation ratio is 0.5 (\ie reduction in 50\% of the cyberattacks).

Equation \ref{eq:ROSI-example} shows the calculation for the example explained above. As it can be seen, the value of the calculated ROSI is 0.8, which means that the solution is not cost-effective, since a ROSI $\geq$ 1 is the goal.

\begin{equation}
\label{eq:ROSI-example}
\begin{aligned}
ROSI = \frac{(ALE \times Mitigation_{Ratio}) - Solution_{Cost}}{Solution_{Cost}}\\
ROSI = \frac{(3 \times 30000) \times 0.5) - 25000}{25000}\\
ROSI = 0.8 \text{ (\ie not cost-effective)}
\end{aligned}
\end{equation}

\section{Related Work}
The solutions surveyed for this work are tools, systems, or software that implement methodologies or techniques to allow users to handle cybersecurity demands in a more intuitive way. All of these solutions discussed in this section provide at least \1 a backend that implements a set of features for cybersecurity planning and investment and \2 a frontend that allows users to interact with the solution to access the features. Therefore, solutions like those discussed below are essential for cybersecurity planning and investment, especially for SMEs that need intuitive and simplified ways to handle their cybersecurity.

An overview and comparison of different solutions discussed within this section is shown in Table \ref{table:solutions}. These solutions are classified according to their categories (\eg risk assessment or cybersecurity investment) and whether they are commercial products or research prototypes. Also, the availability of important features to support cybersecurity planning and investment are analyzed, including the \1 availability of intuitive and user-friendly interfaces to users access the features provided by the solutions, \2 technical dimensions of cybersecurity, such as analysis of vulnerabilities, identification of violations of CIA triad, and integration with robust monitoring solutions, and \3 coverage and analysis of economic aspects of cybersecurity, such as considering financial loss due to cyberattacks and cost-effective decisions for better cybersecurity. 

\begin{table*}[h!]
  \centering
      \caption{Overview and Comparison of Solutions for Cybersecurity Planning and/or Investment}
      \label{table:solutions}
 \begin{tabular}{c|c|c|c|c|c|l} 
 \hline 
 \textbf{\makecell{Solution}} & \textbf{\makecell{Category}} & \textbf{\makecell{Type}} & \textbf{\makecell{User-Friendly \\ Interface}} & \textbf{\makecell{Technical \\ Aspects}} & \textbf{\makecell{Economic \\ Aspects}} & \textbf{\makecell{Characteristics}} \\
 \hline
 \makecell{Cybersecurity \\ Osservatorio\\ Questionnaire \cite{osservatorio}} & \makecell{Risk \\ Assessment} & \makecell{Product} & \makecell{Yes} & Partially & Yes & \makecell*[{{p{3.8cm}}}]{Provides report on expected annual losses.} \\
 \hline
 \makecell{Rea-Guaman \\ et al. \cite{8399252}} & \makecell{Risk \\ Assessment} & \makecell{Research \\ and Prototype} & \makecell{Yes} & Partially & Partially & \makecell*[{{p{3.8cm}}}]{Correlation between  Vulnerabilities and Assets.} \\
 \hline
  \makecell{CERCA \cite{CONCORDIA-D43}} & \makecell{Risk \\ Assessment} & \makecell{Product} & Yes & Yes & Yes & \makecell*[{{p{3.8cm}}}]{Real-time Assessment and SIEM integration.} \\
 \hline
 \makecell{Tracking \\ Vulnerabilities \cite{trackingvul}} & \makecell{Recommender\\ System and Risk \\ Assessment} & \makecell{Research \\ and Prototype} & \makecell{No} & Yes & No & \makecell*[{{p{3.8cm}}}]{NLP and ML techniques is applied to list vulnerabilities in a software inventory.} \\
 \hline
 \makecell{ReCIst \cite{complexis20}} & \makecell{Cybersecurity \\ Investment} & \makecell{Research \\ and Prototype} & Yes & Partially & Yes & \makecell*[{{p{3.8cm}}}]{Quantifies the effects of cybersecurity investment in critical infrastructures.} \\
 \hline
  \makecell{CET \cite{BENZ2020531}} & \makecell{Risk \\ Management} & \makecell{Research \\ and Prototype} & No & Yes & Partially & \makecell*[{{p{3.8cm}}}]{Questionnaire-based tool with 35 questions based on NIST CSF.} \\
 \hline
 \makecell{SERViz \cite{Inan20}} & \makecell{Cybersecurity \\ Planning} & \makecell{Research \\ and Prototype} & \makecell{Yes} & Partially & Partially & \makecell*[{{p{3.8cm}}}]{Protection measures and ROSI integration.} \\
 \hline
 \makecell{CSAT \cite{8993090}} & \makecell{Risk \\ Management} & \makecell{Research \\ and Prototype} & Yes & Yes & No & \makecell*[{{p{3.8cm}}}]{Visual tool that simplifies and automates the application of NIST CSF in companies.} \\
 \hline
 \makecell{Li et al. \cite{whatdata}} & \makecell{Recommender\\ System and \\ Cybersecurity \\ Planning} & \makecell{Research \\ and Prototype} & Yes & Yes & Yes & \makecell*[{{p{3.8cm}}}]{Provides recommendation for data protections based on risk factors and a given budget.} \\
 \hline
  \makecell{MENTOR \cite{MENTOR}} & \makecell{Recommender\\ System for Protections} & \makecell{Research \\ and Prototype} &  Yes & Partially & Partially & \makecell*[{{p{3.8cm}}}]{Provides recommendation for protections based on technical demands and a given budget.} \\
 \hline
 \end{tabular}
\end{table*} 

The Cybersecurity Osservatorio offers a set of cybersecurity services to raise SMEs' awareness of the importance of cybersecurity. One of these services is composed of a cybersecurity self-assessment tool \cite{osservatorio}. The main goal of the tool is to provide a quick and straightforward tool for cyber risk self-assessment. The tool requires two types of input: information about security measures and information about key assets of the enterprise. When all inputs are provided, the tool estimates the expected annual losses for every relevant threat and a total one. The output is to be available when the input information is correctly provided. 

\cite{Inan20} developed a tool named SERViz to support the risk assessment and economic analysis of cybersecurity. By using the tool, decision-makers can configure different parameters related to their business (\eg business sector, operation systems, and most common attacks), analyze the risks, obtain insights about direct and indirect costs (\eg due to business's downtime and reputation loss), and receive recommendations of cost-efficient countermeasures. The tool also relies on ROSI to highlight, besides technical aspects, the better from the economic perspective. However, the tool is still a conceptual prototype, requiring more investigation related to available cybersecurity economics metrics, validation with industry partners, and populated with relevant cybersecurity information to become used in production. 

\cite{trackingvul} proposed a recommender system that tracks and recommends protection against vulnerabilities. The work uses a pipeline composed of Natural Language Processing (NLP), fuzzy matching, and ML. The automation provided by the solution allows a cybersecurity analyst to obtain a list of vulnerabilities that match their software or hardware inventories. The recommender system was tested and compared against a human analyst. During the evaluation, 50 software and 50 hardware inventories with commonly used software, network, and computer hardware components were considered. As a result, for these given datasets, the recommender system saved over 7 hours of work while also providing more accurate results than vulnerability analysis conducted by humans. Although other recommender systems are available in the literature, this is, according to the authors, the first work to address, in an automated way, the problem of the matching of vulnerabilities in private inventories of software and hardware.

A recommender system for data protection was introduced by \cite{whatdata}, which simulates protection options and provides insights into aggregated plans. The system recommends protections for a given data group to achieve a higher risk deduction with a given budget. Also, related risk factors can be visualized in the user interface, allowing for interactions and recommendations according to the user's demands. This kind of system can reduce security analysts' cognitive load and improve the performance of tasks required for efficient data protection decisions. Even though this work can serve as the first step toward data-centric security application, the authors emphasize that evaluations with larger samples are still needed to validate and improve the proposed system. 

Similarly, MENTOR \cite{MENTOR} was also proposed as a recommender system for protections relying on correlation measurements to determine which protections fit better businesses' demands (\eg type of attack, region, and leasing period) and budget available. MENTOR was integrated with different solutions in the cybersecurity planning process, such as the conversational agent for cybersecurity planning proposed in \cite{secbot} and the blockchain-based marketplace for protections introduced in \cite{protectddos}.

In another work based on NIST CSF, the authors proposed a user-interactive cybersecurity tool to simplify and automate the NIST-compliance of companies \cite{8993090}. This work developed a front-end and back-end to provide a robust and user-friendly NIST-compliance guideline tool. For that, features were developed, such as the questionnaires generators based on NIST CSF according to the company being analyzed, a heat map generator to visualize the CIA score, and a database editor for information management. Also, APIs were developed to allow the interaction between the different components and features of the work. The work was validated in a scenario considering e-commerce risk management. However, even simplifying the process by providing Web-based interfaces and other features, applying the NIST CSF remains a challenge for SMEs, since it requires an understanding of cybersecurity-related information, concepts, and interactions.

In \cite{8399252}, it was proposed a new solution for the analysis and risk management. The solution novelty relies on the correlation between vulnerabilities and assets available in the company. It is possible to understand the potential impacts on the assets if a given vulnerability is exploited or an incident happens. The authors argued a gap in the literature that concerns technical and economic impacts, since most of the solutions available for risk management focus on the threats only without understanding the assets and their possible economic impacts.

A tool named ReCIs was introduced in \cite{complexis20}. The tool applies the Return on Cybersecurity Investment (ROCI) model, also proposed by the authors of the work, to quantify the effect of cybersecurity investment on critical infrastructure. In the ROCI model, the ultimate return value to determine if protection is cost-efficient is calculated as the annual difference between costs associated with cyberattacks minus the costs of those same attacks, now mitigated by a cybersecurity solution. This metric looks very similar to the ROSI, one of the metrics covered by this work. By using the ROCI model and recommender systems, the ReCIs tool can provide financial cost overviews. Also, it helps during the decision process of selecting a cybersecurity solution for critical infrastructure. This work was one of the first cybersecurity investment approaches to quantify a return on investment for the critical infrastructure sector

As an example of effort from the industry, there are dedicated efforts to developing a tool for real-time risk assessment called CERCA \cite{booklet-ATOS}. The tool receives input data from various sources that can inform about changes in the target system (\eg new threats, new target nodes, new vulnerabilities, alarms from Security Information and Event Management, and Intrusion Detection System tools).

Based on the literature review, the cybersecurity field is receiving much attention on different fronts, from mitigating cyberattacks to planning and investing in defining cybersecurity strategies. However, there are opportunities for multidisciplinary approaches to address cybersecurity gaps focusing on more efficient, economically viable, and sustainable strategies.

\section{The SECAdvisor Tool}
The SECAdvisor is a solution proposed to support the definition of budget, requirements, and information during the cybersecurity planning of companies. To the best of the authors' knowledge, SECAdvisor is the first solution of its kind to integrate different cybersecurity economic models in cybersecurity planning in a straightforward and extensible way. The solution was designed and developed to support companies but also to support training and educational activities with people interested in cybersecurity planning, such as consultants, academic students, and researchers.

\begin{figure}[h!]
\centering
\includegraphics[width=\linewidth]{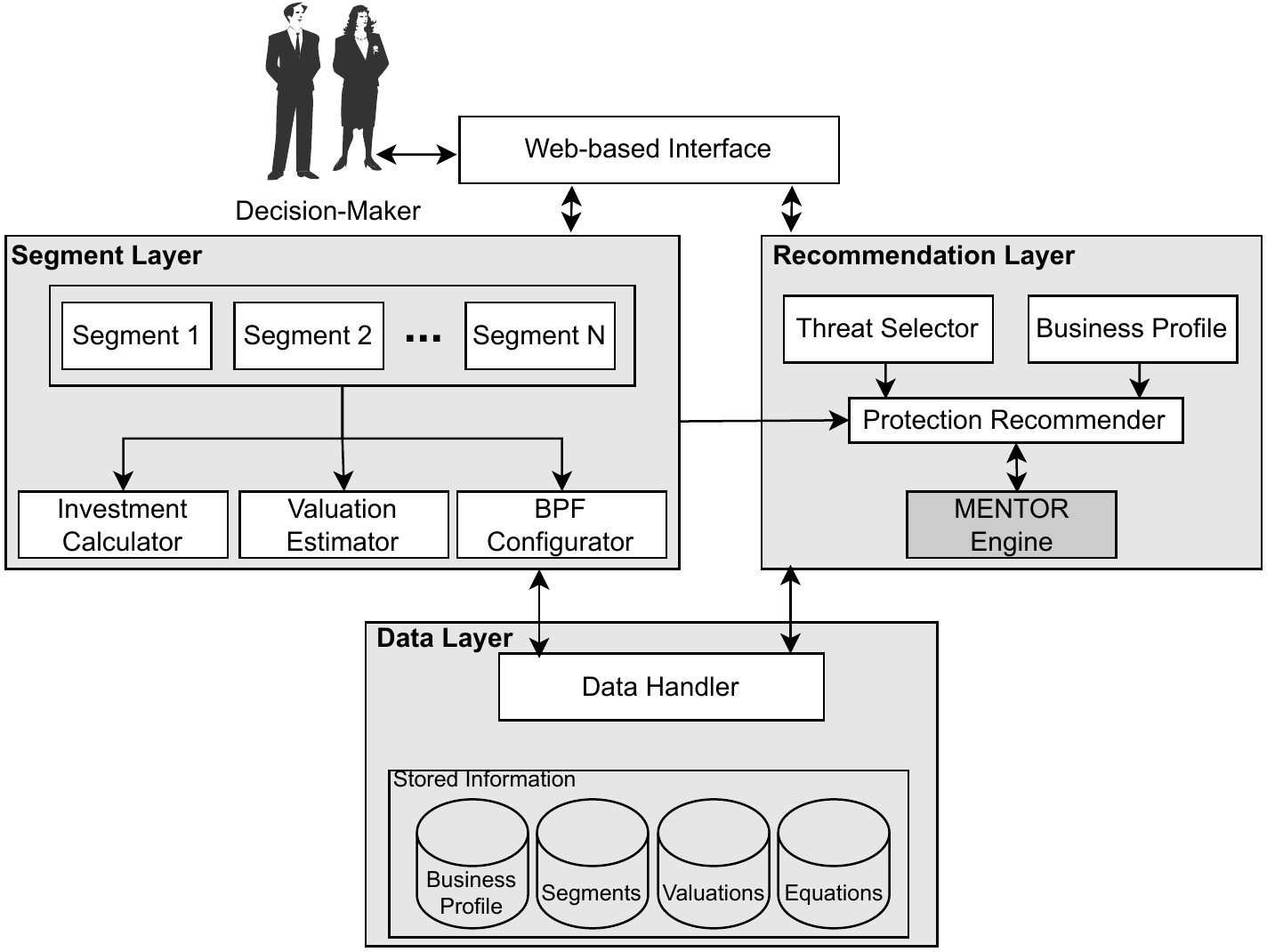}
\caption{Conceptual Architecture of the SECAdvisor}
\label{fig:SECAdvisor}
\end{figure}

Figure~\ref{fig:SECAdvisor} gives an overview of the conceptual architecture of SECAdvisor, including three different application layers and their relationships. The flow starts with the decision-maker (\ie user) accessing the Web-based Interface of SECAdvisor and defining the business profile representing the company they want to conduct the calculations. To create such a profile, the user must submit key information about the company to SECAdvisor, such as the revenue, sector, and number of employees. Next, the Segment Layer is in charge of \1 managing the different segments within the company, \2 estimating how valuable each segment is (\eg based on the critical data available and specific parameters for a given segment), \3 calculating and comparing the optimal investment per segment, and \4 configuring the Breach Probability Function (BPF) according to the needs. Finally, the \textit{Recommendation Layer} allows for the selection of specific threats and, based on the optimal calculation provided by the \textit{Segment Layer}, can determine which protections are suitable for the company in terms of fitting the optimal investment, budget available, and demands to mitigate/avoid a selected~threat.

The \textit{Recommendation Layer} prepares all information required and makes requests to the recommendation engine implemented by MENTOR \cite{MENTOR}. After a list of protections is recommended for the company, the SECAdvisor also calculates the ROSI metric for each protection, since it can support cost-efficient investments by comparing different recommended protections. The \textit{Data Layer} is also implemented by SECAdvisor to store all relevant data (\eg information regarding the business, segments, and knowledge used for the segments estimations). Besides that, all configurations needed for the GL model and for the customization of the SECAdvisor are stored in a database. Therefore, although predefined equations and configurations are placed, the SECAdvisor can be extended and adapted by changing key fields in the database.

\subsection{Segments and Value Estimation}
\label{sec:segments-estimate}
Determination of the segments and their values is critical for the optimal calculation of investments and the recommendation of protections according to specific demands \cite{gordon-segmentation}. A segment represents a technical business area of a company. Thus, the optimal investment amount should be calculated per segment, since a specific segment might be directly related to the potential benefits of cybersecurity investments. The following information is required to determine a new~segment:

\begin{itemize}
\item \textbf{Segment Name:} The parameter represents the segment's name, which can be freely chosen by the company (\ie user).
\item \textbf{Segment Type:} The type of segment is used to suggest suitable cybersecurity threats and to simplify the monetary valuation of the segment. SECAdvisor allows for the selection of different pre-defined segments, such as \textit{Web Server}, \textit{Network}, or \textit{Database} segments. 
\item \textbf{Value:} In order to calculate the optimal cybersecurity investment level, the monetary value (US\$) of the segment is needed. Since it is often difficult to determine this value, the SECAdvisor provides assistance for the valuation of the segment based on publicly available reports and data.
\item \textbf{Risk:} The \textit{Risk} parameter describes the probability of a cyberattack. The user is allowed to specify a number between 0\% and 100\%. This parameter is needed to determine the optimal investment.
\item \textbf{Vulnerability:} This parameter is also needed to calculate the optimal cybersecurity investment. It describes the probability that a cybersecurity attack on the segment will be successful. Values between 0\% and 100\% are allowed.
\end{itemize}

Next, the value of the segment must be estimated. However, it is not a trivial task for a user to determine the monetary value of the segment, such as how much a database is worthy for the business or networking infrastructure. Therefore, the SECAdvisor provides aid to facilitate this decision. The system allows the user to enter parameters tailored to the segment, which are evaluated based on previous knowledge populated in the database (\eg values based on estimations made by reports, research, or shared by partners). Thus, the user can receive a suggestion for the segment's value, which he/she can use as it is or adapt according to his/her view.

\subsection{Investment Calculation} % Check
The SECAdvisor calculates the optimal cybersecurity investment based on an extension of the GL model proposed by~\cite{gordon-segmentation}. This extension combines the GL with the idea of information segmentation. An important factor for the investment calculation is the breach probability function. It is denoted as \textit{S(z,v)}, where \textit{z} describes the monetary investment and \textit{v} the vulnerability of the segment. The breach probability function describes the productivity of the investment, which first increases and then decreases after a certain point. Each additional investment is higher than the resulting benefit from this point on. The steps and definitions described in Section II are used to showcase the application of the GL model within SECAdvisor.

To calculate the optimal investment in cybersecurity, the SECAdvisor uses the breach probability function defined in Equation \ref{breach-function}. Thanks to this GL model extension, the SECAdvisor calculates the optimal investment level for each segment. In addition, the monetary advantage of information segmentation is also illustrated in the application. Note that these equations are fully extracted from the original work that extended the GL model to support information segmentation \cite{gordon-segmentation}. Therefore, it tries to generalize the BPFs to cover hypothetical scenarios anchored by some assumptions related to the reality of cybersecurity today. However, this is not true for any company that wants to invest in cybersecurity. Thus, for an accurate optimal investment calculation, the BPF has to be defined according to the reality and demands of a given company or sector.

\begin{equation}
    \label{breach-function}
    S_{i}(z_{i},v_{i})=\dfrac{v_{i}}{1+\dfrac{1}{L \times 0.001}\dfrac{z}{\dfrac{L_{i}}{L}}}
\end{equation}

To determine the cost-effectiveness of a cybersecurity investment, the SECAdvisor then uses the ROSI metric. This metric is used because cybersecurity investments do not bring a direct profit but reduce potential damage. During the evaluation of cybersecurity investments, the focus is on assessing how much potential loss can be prevented by an investment. Therefore, the monetary value of the investment must be compared with the monetary value of the risk reduction.

\begin{figure*}[ht]
\centering
\includegraphics[width=\textwidth]{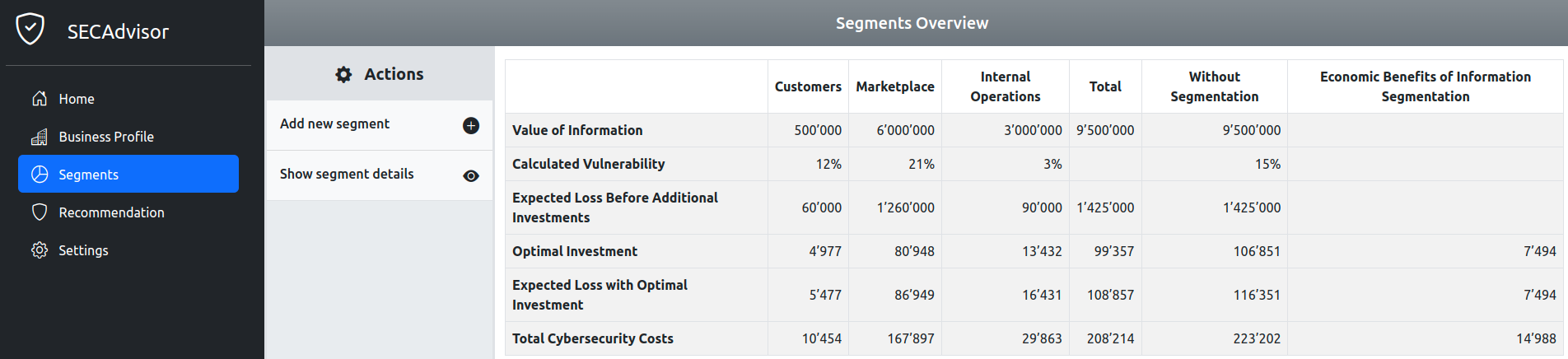}
\caption{Dashboard of SECAdvisor with the Optimal Investments per Segment Calculated}
\label{fig:secadvisor-calculation}
\end{figure*}

\subsection{SECAdvisor's Implementation} % TBD
The SECAdvisor was mainly implemented using \textit{AngularJS} and \textit{NestJS} frameworks. The database is the \textit{MongoDB}, a document-oriented NoSQL database management system. The data used for the SECAdvisor prototype is stored on \textit{MongoDB Atlas}, a flexible and scalable cloud database service. The database was connected using the \textit{Mongoose}, an Object Data Modeling (ODM) library for Node.js. The calculation of the optimal investment level is a core competence of the application. For that, the library \textit{nerdamer} was used to enable calculation operations that JavaScript does not provide by default. Finally, the integration with the recommender system so-called MENTOR was performed by making calls for a RESTful API implemented by MENTOR. The source code and a full-fledged prototype of SECAdvisor are publicly available at \cite{secadvisor-code}, with an official instance running at \url{https://secadvisor.figueredofranco.com}.

The tool is built as docker containers from source code that includes all dependencies as locked dependencies. While this ensures that the tool can always be built, it does not yet prevent upstream dependency attacks that could be used to steal business critical data. Next-generation distribution could include prebuilt images to simplify the deployment process even more while retaining the security aspect through reproducibility.

\begin{figure}[ht]
\centering
\includegraphics[width=0.75\linewidth]{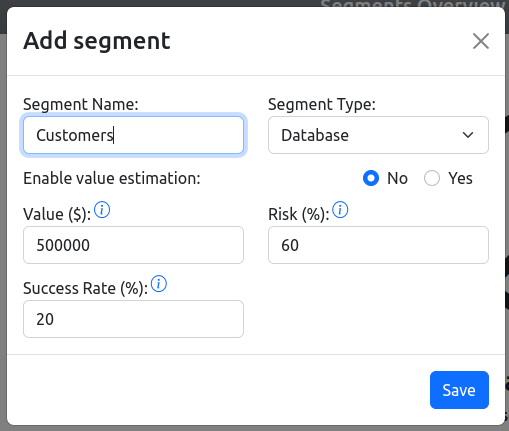}
\caption{Definition of Segments Using the SECAdvisor Interface}
\label{fig:segments-add}
\end{figure}

For the tool's usage, as the first step after the login, the user has to add his/her business profile and the segments that compose the business under analysis. The user can use the SECAdvisor interface to add each segment required for the optimal investment calculation. Figure \ref{fig:segments-add} shows the interface for adding one segment. A database segment (\ie segment type) is selected, which requires the user to fulfill the information regarding the records stored in such a database (\eg number of records with sensitive and anonymized data). If this information is available, the value estimation can be performed automatically; otherwise, the segment's value must be defined manually. Also, the risk of an attack happening in this segment has to be defined together with the likelihood of a successful attack.

% Write about the Web-based interface now

After having the segments determined (\ie value, risk, and vulnerability of a segment), optimal investments can be calculated by applying the GL model. The equations are used as defined by the database (\eg BPF and additional calculations). Figure \ref{fig:secadvisor-calculation} shows the calculations made for each segment added to the SECAdvisor. In this example, three segments are available: Customers Database, Marketplace server, and Internal Operations network. An overview of information is available in the table generated by SECAdvisor, and the optimal investment is defined.

\begin{figure}[ht]
\centering
\includegraphics[width=\linewidth]{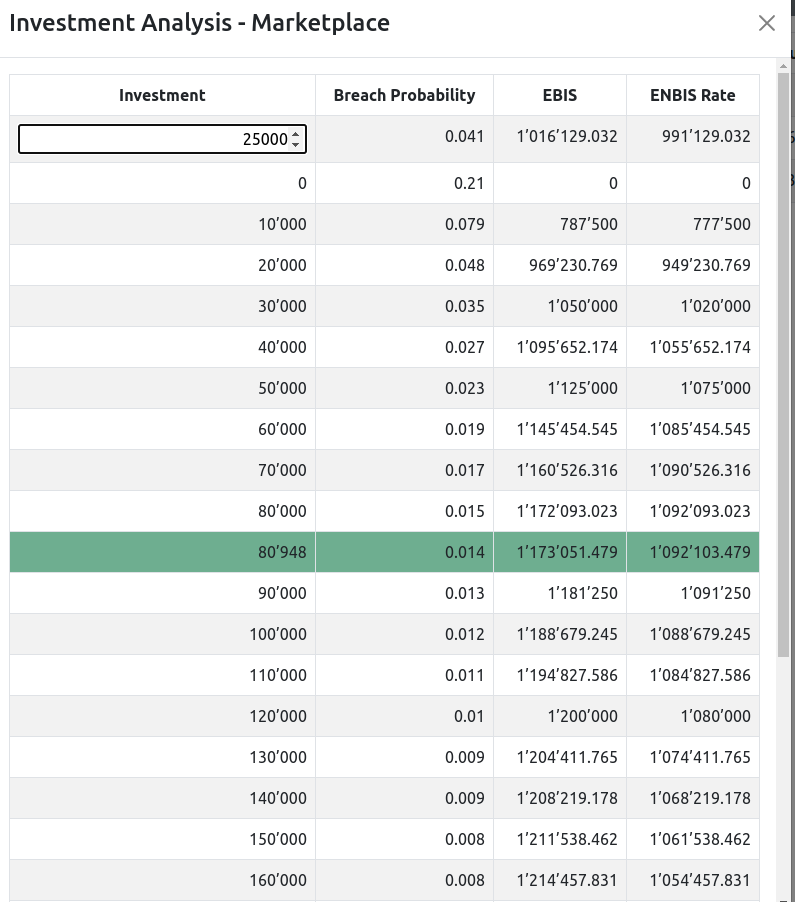}
\caption{Interface for Customization and Zoom-In on the Optimal Investment Calculation}
\label{fig:secadvisor-analysis}
\end{figure}

Furthermore, to explain the calculations in a detailed manner, SECAdvisor provides a feature that shows the different values computed until finding the optimal investment, including values higher and lower than the final value. Figure~\ref{fig:secadvisor-analysis} depicts the ENBIS rate for each value calculated until finding the optimal investment (highlighted in the green row). Also, the user can customize his/her own investment to check if there is a positive ENBIS. Such a feature helps to understand if the current investment decisions are efficient compared to the optimal investment.

The BPF can also be configured according to the preference of the company. Figure \ref{fig:secadvisor-calculation} shows this configuration feature. SECAdvisor uses, as default, the BPF introduced in~\cite{gordon-segmentation}, but allows for \1 changing the weight of each variable part of the equation (\ie basic configuration) or \2 defining a complete new BPF (\ie advanced configuration). Moreover, it is possible to compare different BPFs with the original one provided by the GL model. Thus, users can adjust the BPF according to their companies' reality to calculate the optimal~investment.

\begin{figure}[ht]
\centering
\includegraphics[width=\linewidth]{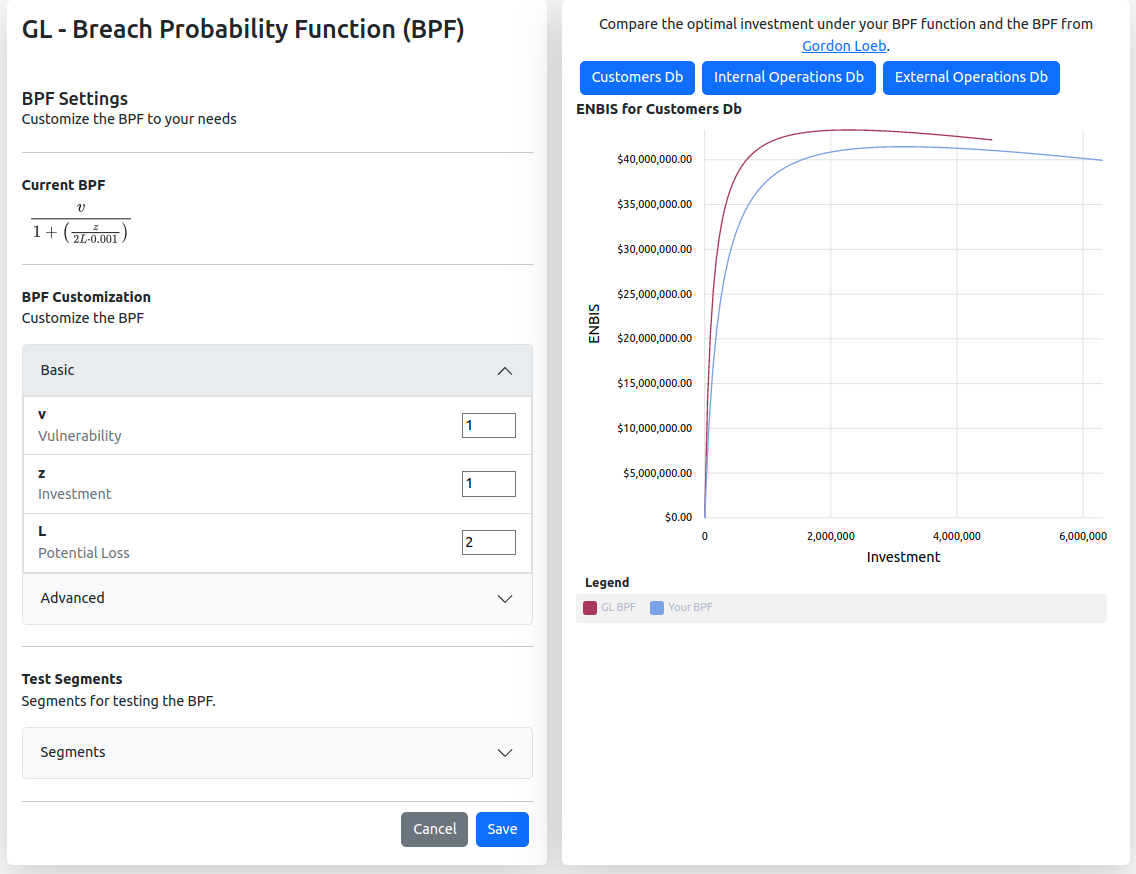}
\caption{Interface for Customization of BPF using SECAdvisor}
\label{fig:secadvisor-calculation}
\end{figure}

After the optimal investment calculation, the user can use this information as input for the next steps of planning and investment, taking it as a reference value for each segment. For instance, this value can determine the maximum budget to spend with protections for a specific segment. With this amount at hand, the user can then go for the \textit{Recommendation tab}, which allows obtaining recommendations of protections based on the MENTOR engine. Besides recommending protections that fit the budget (\ie optimal investment) and customized demands, the SECAdvisor calculates the ROSI metric by just clicking right below one suggested protection. 

\begin{figure}[h!]
\centering
\includegraphics[width=0.75\linewidth]{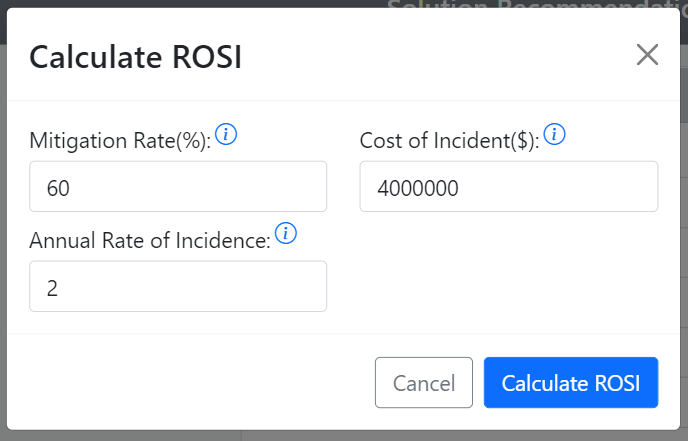}
\caption{Input Parameters for ROSI Calculation for Protections using SECAdvisor}
\label{fig:secadvisor-recommendation}
\end{figure}

The ROSI calculation can be done for each protection recommend, which requires the user to provide the mitigation rate, the incident cost, and the annual incidence rate for each protection, as shown in Figure \ref{fig:secadvisor-recommendation}. According to the segment definition, these values are already received from the MENTOR recommendation engine and provided by SECAdvisor. However, this can be manually edited by the user if needed. After receiving the recommendation and the value for ROSI (\ie a ROSI higher than one means cost-efficient protection), the user can decide which protection to acquire to achieve sufficient protection while investing only the optimal amount in cybersecurity.

\section{Evaluations}
The evaluation conducted on SECAdvisor focuses on \1 the usability and benefits of SECAdvisor as well as \2 highlights examples of successful practical activities conducted using SECAdvisor for educational purposes in European cybersecurity~courses. 

It is important to note that the evaluation of GL is explicitly out of the scope since it is already extensively evaluated in the literature, such as in \cite{gordon-37, gordon-calibrate, gordon-insurance, gordon-segmentation}. Therefore, this section focuses on evaluating the capacity of the SECAdvisor tool to achieve the correct values of optimal investments by applying the GL model. Also, the evaluation of the recommendation process was already conducted in additional research available at \cite{MENTOR}.

\subsection{Usability}

For the usability evaluation, a survey was conducted on the platform's usability and on intuitiveness of the platform for its use for cybersecurity investment calculations. Real-world users were invited to use the platform and fulfill a given set of tasks and rate its usability on a System Usability Scale (SUS) questionnaire. The usability evaluation used a methodology similar to the ones applied in~\cite{MF-ma}.

Thirteen people participated in the evaluation of the SECAdvisor. All users were able to create an account on the platform and log in to the tool. All participants were able to use the tool successfully and did not face any technical problems. Survey participants all ranged in the age group of 20--49, with the majority of respondents (7) from 20 to 29. The survey participants covered various educational levels, from bachelor's to doctoral degrees. 

Users were asked to create an account and configure SECAdvisor with essential information for three segments: Web Server, Database, and Network. After this initial configuration, a set of tasks was defined and conducted to evaluate different features implemented in the SECAdvisor tool. They were tailored to validate if most users can find correct results by applying cybersecurity economics principles in an intuitive and user-friendly way. Table~\ref{tab:tasks} summarizes the tasks conducted and their success rate. A task was considered successfully solved if the answer matched the correct answer. For all tasks, the majority of the participants were able to solve the task successfully and provide the correct answer.

\begin{table*}[h]
\centering
\caption{Tasks Performed by Participants using the SECAdvisor}
\label{tab:tasks}
\begin{tabular}{c|c|c|c} 
 \hline 
 \textbf{Task} & \textbf{Question} & \textbf{Answer} & \textbf{Success Rate}   \\ \hline
\makecell{1} & \makecell{What is the vulnerability of the Database?} & \makecell{8\%} & \makecell{92\%} \\ \hline
\makecell{2} & \makecell{What is the yearly expected loss of the Database \\ if there are no additional investments in cybersecurity?} & \makecell{\$ 24,576} & \makecell{100\%} \\ \hline
\makecell{3} & \makecell{After adding all the segments in the tool, \\ how much is the economic benefits between the \\ investment using information segmentation and without \\ information segmentation considering the optimal investment?} & \makecell{\$ 1,852} & \makecell{77\%} \\ \hline
\makecell{4} & \makecell{How much is the total costs of cybersecurity \\ for all of the segments?} & \makecell{\$ 41,079} & \makecell{92\%} \\ \hline
\makecell{5} & \makecell{Which recommendation provides the highest ROSI \\ for the Network segment?} & \makecell{Portwell} & \makecell{92\%} \\ \hline
\makecell{6} & \makecell{What is the optimal investment for the Database segment \\ after adjusting for 1.5 the weight of the vulnerability (v)\\ on the breach probability function?} & \makecell{\$ 3,058} & \makecell{69.2\%} \\ \hline
 \end{tabular}
\end{table*}

Task 1 had an excellent success rate of 92\%. The only person who was unsuccessful in this task was able to answer all of the other correctly, which could mean that it was simply an error filling out the form or similar. Every participant was able to solve Task 2 successfully. This is not surprising given the results for Task 1, as this is on the same page and clearly described in the task description. Task 3 had a success rate of 77.0\%, which is still satisfactory but lower than Task 1 and 2. This could be explained by the low number of people who indicated familiarity with the GL model's information segmentation concept. As such, finding the benefit of segmentation might have proved harder because of the need for more background knowledge. It could also be influenced by the fact that this question addresses the overall numbers and not the numbers of the individual segments. Task 4 was again solved very successfully, with only a single participant unable to complete the task successfully. It should be further mentioned that the person who selected the wrong answer selected the right amount for the total cost of cybersecurity but chose the wrong parameters used for the calculation.

For Task 5, the participants had to leave the segment overview table and go into the recommendation part. Most participants solved the task successfully, with only one entering the wrong solution. This should strengthen the statement that many participants know about ROSI. This means that users were able to navigate to the recommendation page, input the given data for the recommendations, and then compare the ROSI values. This also underlines the tool's usefulness since most participants successfully got a recommendation for a security product with little to no previous experience with the tool.

Task 6 had the lowest success rate, with 69.2\% being able to adjust the breach probability function and obtain the correct result. This is not a surprising finding because it was the most advanced task and as most users did not have a background in breach probability functions. Still, this comparatively low success rate should not be faced too harshly, mainly because it is not necessary always to input a custom breach probability function while using the tool, as GL already provides a reasonable, widely accepted, and well-researched default function \cite{skeoch_expanding_2022, farrow_cybersecurity_2016, naldi_calibration_2017, gordon_investing_2016, gordon_information_2021}. 

The overall success rate is high, with 87.2\% of correct answers. This means that most participants in the evaluation were able to use the tool properly and use its support to solve the tasks successfully. All evaluation participants had a technical background, were expected to be above average in technical skills, and were more likely to figure out how the system works successfully.

Finally, a final score was calculated according to the SUS~\cite{SUS}. The SUS is not standardized according to any significant standard but is often used in academia to test the usability of software systems. The SUS was employed because it is a \textit{de facto} standard, is simple, and can be used with small sample sizes. Participants had to rate every question on a 5-point scale from 1 (strongly disagree) to 5 (strongly agree). The following ten statements (S) were rated by the participants of the SUS questionnaire, right after conducting the tasks highlighted in Table \ref{tab:tasks}.% of all participants:

\begin{itemize}
%\begin{enumerate}
    \item[\textbf{S1}] I think that I would like to use this system frequently.
    \item[\textbf{S2}] I found the system unnecessarily complex.
    \item[\textbf{S3}] I thought the system was easy to use.
    \item[\textbf{S4}] I think that I would need the support of a technical person to be able to use this system.
    \item[\textbf{S5}] I found the various functions in this system were well integrated.
    \item[\textbf{S6}] I thought there was too much inconsistency in this system.
    \item[\textbf{S7}] I would imagine that most people would learn to use this system very quickly. 
    \item[\textbf{S8}] I found the system very cumbersome to use.
    \item[\textbf{S9}] I felt very confident using the system.
    \item[\textbf{S10}] I needed to learn a lot of things before I could get going with this system.
%\end{enumerate}
\end{itemize}

All ten questions regarding the system usability are awarded a score from 0-4, with 0 being the most negative and 4 being the most positive. These are then summed up over the ten questions, resulting in a value from 0 to 40, which has to be multiplied by 2.5 to get a SUS score on the scale from 0 to 100. It should be noted that this is not a percentage but rather a point on a 0 to 100 scale, which a value above 70 is considered satisfactory. The overall SUS score was 82.1, within the standard deviation from the adjective excellent. Therefore, the usability of SECAdvisor is very good and provides essential features for a user-friendly and intuitive application of cybersecurity economic metrics.

There were three outliers in the SUS scores of the participants. The first did not send any additional feedback that might help explain the rating but indicated that their area of expertise is \textit{Informatics}, the most general value given there. The second mentioned that their area of expertise is \textit{Cybersecurity and Risk Management} which explains the very high understanding of the tool. This also shows that the tool is not void of profiting from specific domain knowledge, and users with experience in the cybersecurity field might be able to use the tool more easily. Finally, the third, with the lowest SUS score, mentioned that they  had trouble understanding what to adjust in the system without step-by-step guidelines. 

\subsection{Real-World Practical Activities}
A set of practical educational activities was conducted using SECAdvisor. This allowed hundreds of people and participants to get contact with cybersecurity economics concepts for the first time or even -- when the knowledge was existent -- understand scenarios where it can be applied usefully. 

The SECAdvisor was already part of the first four editions (from 2020 to 2022) of the course "Becoming a Cybersecurity Consultant"~\cite{CONCORDIA-Course}, which is part of a certification scheme that supports people in preparing for a career as a cybersecurity consultant. The course is an initiative of the H2020 CONCORDIA project and had 360 participants enrolled (at the time of writing). SECAdvisor was used for a 90 minutes practical exercise conducted after the theory on cybersecurity planning with an economic bias. Around 120 participants attended the four practical sessions. The exercises supported by SECAdvisor included the \1 calculation of optimal investments in cybersecurity using the GL-model, \2 identification of protections candidate that fits specific companies' demands, and \3 selection of cost-effective protections from a list of candidates using automated calculation of the ROSI metric. All participants could apply the concepts, and the feedback was mainly positive, highlighting SECAdvisor as the first tool of its kind that provides many benefits for both practical application in industry and education purposes.

Also, SECAdvisor was used for practical exercises during a cybersecurity lecture for 20 Master's students at the University of Zurich UZH, Switzerland. This helped them understand practical applications of cybersecurity concepts and conduct planning tasks. Finally, SECAdvisor was part of a 180 minutes tutorial on the European Network for Cybersecurity (NeCS) PhD School 2023. Around 30 participants, all at the doctoral level and with experience in cybersecurity, had the opportunity to interact with SECAdvisor to conduct a set of five practical tasks for cybersecurity planning. Besides the exercises mentioned above, the tutorial also included the definition of customized security breach probability functions and comparing the current cybersecurity investment budget against the optimal investment.
\\ \\
\section{Conclusions and Future Work}
\label{sec:conclusions}
Solutions like SECAdvisor can benefit SMEs (and also large companies) around the globe in better planning and investment decisions in cybersecurity while supporting the analysis of possible financial losses due to a successful cyberattack. Besides cybersecurity solutions, key investments must be made to increase cybersecurity staff and promote cybersecurity awareness for their general employees. Therefore, companies must ensure they can detect and mitigate cyberattacks effectively, using a clear cybersecurity strategy tailored to the company's reality, thus, targeting personnel culture, size, sector, and budget, while covering all relevant facets of~cybersecurity. 

SECAdvisor was designed and implemented to apply well-known cybersecurity economic metrics (\eg Gordon-Loeb and ROSI) intuitively and straightforwardly. For that, SECAdvisor relies on well-investigated methodologies \cite{CyberTEA, CyberTEA-Paper} to focus on essential phases of the cybersecurity planning, including the \1 understanding of business profile, \2 definition of information segments and their associated risks, \3 calculation of optimal investments in cybersecurity, and \4 recommendation of cost-efficient cybersecurity solutions to address all companies' demands for an efficient cybersecurity~strategy.

The evaluations and activities performed with several real-world users prove the benefits and feasibility of SECAdvisor for disseminating and applying cybersecurity economic concepts for different stakeholders (\eg educators, consultants, security experts, and researchers). The tasks performed during the usability evaluation provide a high task success rate when used by people with technical knowledge, and even advanced features can be employed successfully. The tool strives to fulfill different criteria by providing relevant features, such as providing user-friendly interaction for non-technical users and simplifying the process for calculating the optimal investment in cybersecurity. 

Future work includes the evaluation of the tool with the industry using data for real vulnerabilities, threats, and controls. Also, Monte Carlo simulations can be used to assess the decisions of the tool. The efficacy of protection measures and their overlapping is still an open challenge that have to be investigate further. Besides, collaborative approaches for more precise calculations regarding risks and investments are mapped as extensions for the tool. In addition, the customization of the BPF can be enhanced so that it is more intuitive and automated based on different profiles of companies and sectors. Finally, investigations on novel cybersecurity economic models can be conducted, including the applications of artificial intelligence techniques to infer correctly the information needed (\eg costs, risks, and assets valuation) for a proper calculation of cybersecurity~investments.

\section*{Acknowledgement}
This work was supported partially by \textit{(a)} the University of Z\"urich UZH, Switzerland and \textit{(b)} the European Union's Horizon 2020 Research and Innovation Program under Grant Agreement No. 830927, the CONCORDIA project.
\newpage
\bibliographystyle{IEEEtran}
\bibliography{main.bib}
\noindent \small{\\All links provided above were last accessed on April, 2023.}

\input{authors.tex}

\end{document}

%% file: authors.tex
\begin{IEEEbiography}[{\includegraphics[width=1in,height=1.25in,clip,keepaspectratio]{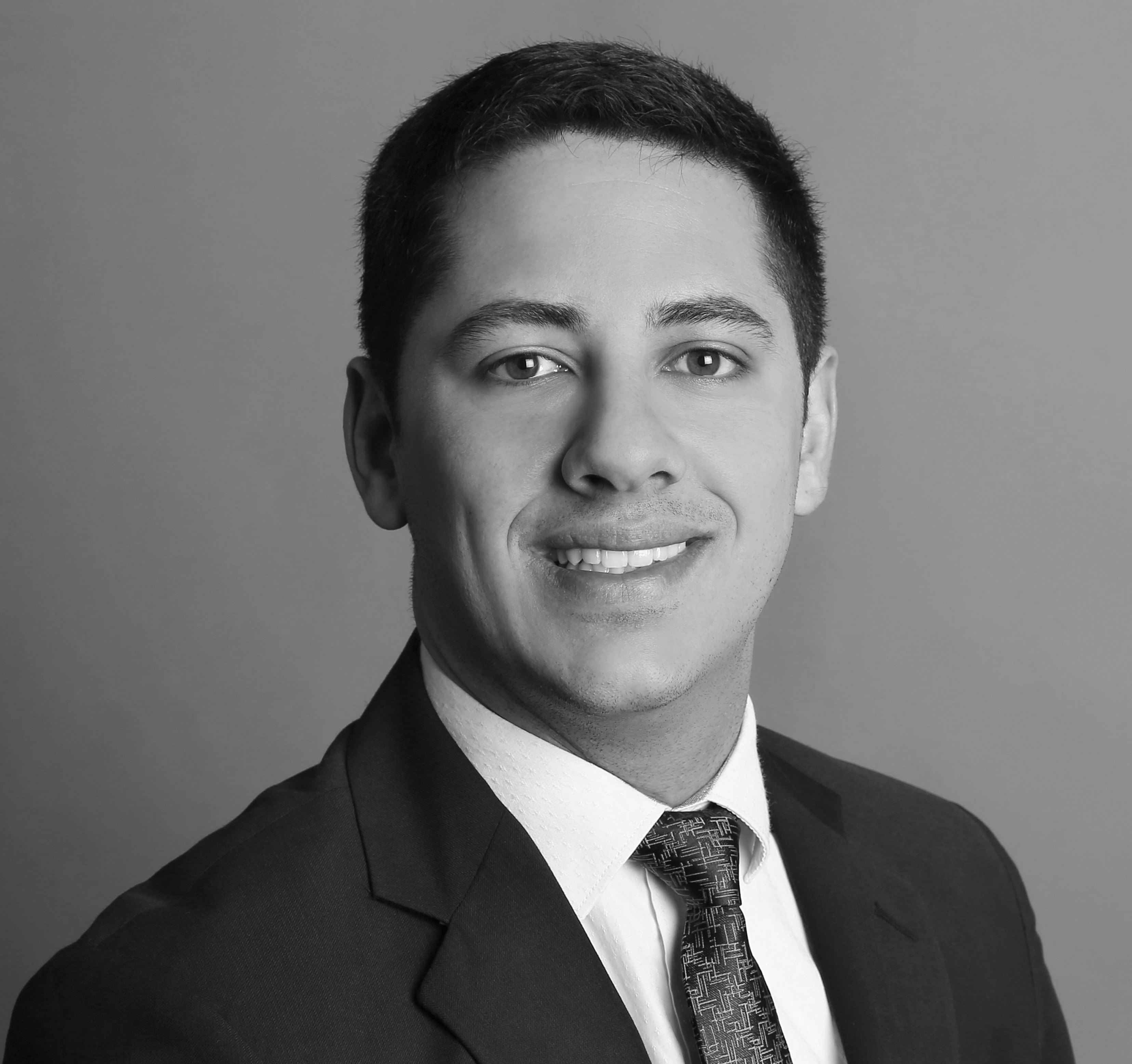}}]{Muriel F. Franco}
is a Researcher in Computer Science at University of Z{\"u}rich UZH, Switzerland, within the Communication Systems Group CSG of the Department of Informatics IfI. Muriel participated in different multidisciplinary projects, participating and driving the work of within a team of networking, security, and economic researchers, several research papers and patents co-authored. Muriel holds an PhD (Summa cum laude) from 2023 in Computer Science from University of Zurich UZH, Switzerland, MBA from 2022 in Project Management from University of São Paulo (USP), Brazil, MSc from 2017 in Computer Science from the Federal University of the Rio Grande do Sul (UFRGS), Brazil, and obtained a BSc from 2014 in Computer Science from the Federal University of Pelotas (UFPEL), Brazil. His research interests include cybersecurity, network management, information visualization, and communication systems. \\
\vspace{-3cm}
\end{IEEEbiography}
\begin{IEEEbiography}[{\includegraphics[width=1in,height=1.25in,clip,keepaspectratio]{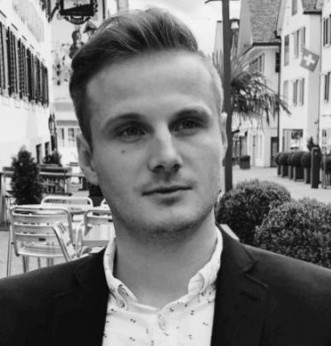}}]{Christian Omlin}
is pursuing his MSc in Computer Science with major in Information Management at the University of Z{\"u}rich UZH, Switzerland. Christian holds an BSc degree from the University of Z{\"u}rich UZH, Switzerland. His research interests include computer networks, cybersecurity economics and information visualization.\\ \\ \\
\end{IEEEbiography}
\vspace{-3.5cm}
\begin{IEEEbiography}[{\includegraphics[width=1in,height=1.25in,clip,keepaspectratio]{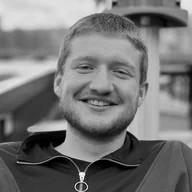}}]{Oliver Kamer}
is pursuing his MSc in Computer Science with major in Software Systems at the University of Z{\"u}rich UZH, Switzerland. Oliver holds an BSc degree from the University of Z{\"u}rich UZH, Switzerland. His research interests include computer networks, cybersecurity economics and information visualization. \\ \\ \\ \\
\end{IEEEbiography}
\vspace{-3.5cm}
\begin{IEEEbiography}[{\includegraphics[width=1in,height=1.25in,clip,keepaspectratio]{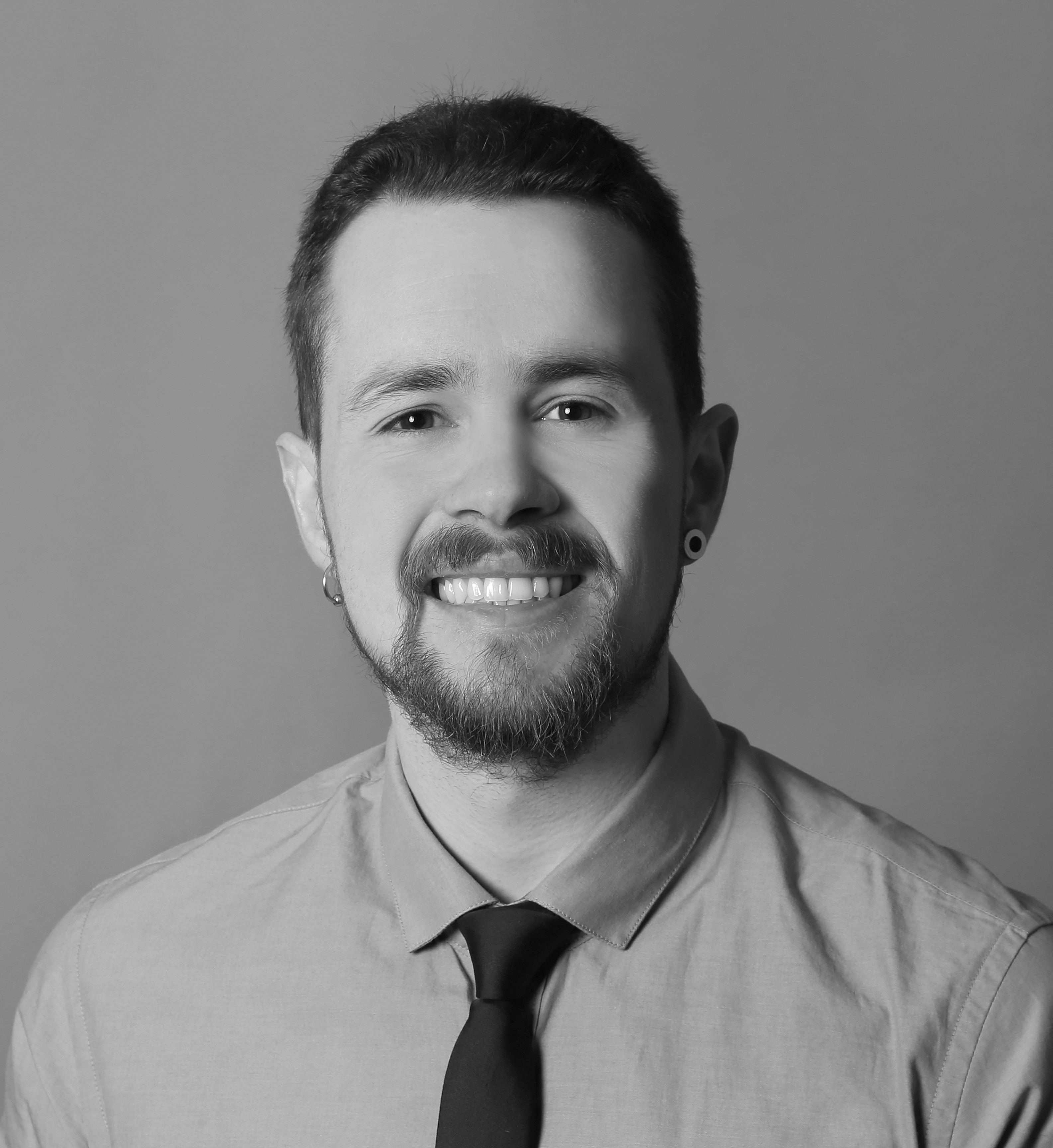}}]{Eder J. Scheid} received his doctoral degree in 2022 from the University of Z{\"u}rich UZH, Switzerland, within the Communication Systems Group CSG of the Department of Informatics IfI. Eder holds an MSc degree in Computer Science from the Federal University of the Rio Grande do Sul (UFRGS), Brazil, which he obtained in 2017. Eder focuses his research on blockchains, smart contracts, policy- and intent-based network management, and NFV.
\end{IEEEbiography}
\vspace{-2.5cm}
\begin{IEEEbiography}[{\includegraphics[width=1in,height=1.25in,clip,keepaspectratio]{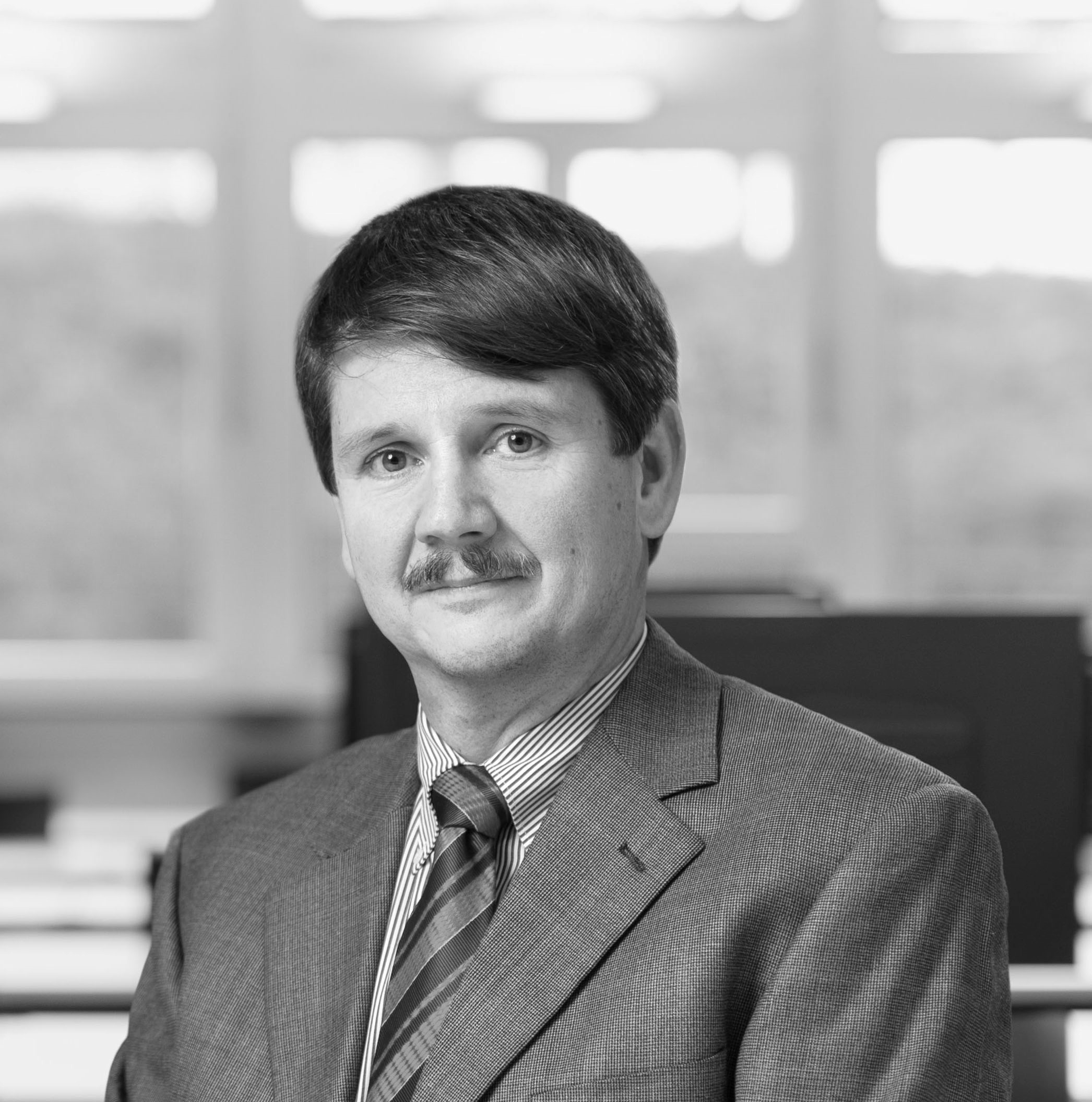}}]{Burkhard Stiller}
received the Informatik-Diplom (MSc) in Computer Science and the Dr. rer.-nat. (PhD) degree from the University of Karlsruhe, Germany, in 1990 and 1994, respectively. In his research career he was with the Computer Lab, University of Cambridge, U.K. (1994-1995), ETH Z{\"u}rich, Switzerland (1995-2004), and the University of Federal Armed Forces Munich, Germany (2002-2004). Since 2004 he chairs the Communication Systems Group CSG, Department of Informatics IfI, at the University of Z{\"u}rich UZH, Switzerland. Besides being a member of the editorial board of the IEEE Transactions on Network and Service Management, Springer’s Journal of Network and Systems Management, and the KICS’ Journal of Communications and Networks, Burkhard is the past Editor-in-Chief of Elsevier’s Computer Networks journal. His main research interests are published in well over 300 research papers and include systems with a fully decentralized control (Blockchains, clouds, peer-to-peer), network and service management (economic management), Internet-of-Things (security of constrained devices, LoRa), and telecommunication economics (charging and accounting).
\end{IEEEbiography}